\documentclass{article}

\usepackage[backend=biber,style=authoryear,uniquename=false,maxcitenames=2,maxbibnames=99]{biblatex}
\usepackage{svg}
\usepackage{amsmath} 
\usepackage{float}
\usepackage{tabularx}
\usepackage{multirow}
\usepackage{todonotes}
\usepackage{arxiv}
\usepackage{lipsum}
\usepackage[utf8]{inputenc} 
\usepackage[T1]{fontenc}    
\usepackage{hyperref}       
\usepackage{url}            
\usepackage{booktabs}       
\usepackage{amsfonts}       
\usepackage{nicefrac}       
\usepackage{microtype}      
\usepackage{lipsum}
\usepackage{graphicx}
\graphicspath{ {./images/} }

\title{Sketch2Simulation: Automating Flowsheet Generation via Multi Agent Large Language Models}

\author{
 Abdullah Bahamdan \\
  Sargent Centre for Process Systems Engineering\\
  Imperial College London\\
  London, SW7 2AZ, United Kingdom \\
 \texttt{a.bahamdan22@imperial.ac.uk} \\
   \And
 Emma Pajak \\
  Sargent Centre for Process Systems Engineering\\
  Imperial College London\\
  London, SW7 2AZ, United Kingdom \\
 \texttt{emma.pajak19@imperial.ac.uk} \\
  \And
 John D. Hedengren \\
  Department of Chemical Engineering\\
  Brigham Young University\\
  Provo, Utah 84602, United States \\
 \texttt{ john.hedengren@byu.edu} \\
  \And
 Antonio del Rio Chanona \\
  Sargent Centre for Process Systems Engineering\\
  Imperial College London\\
  London, SW7 2AZ, United Kingdom\\
 \texttt{a.del-rio-chanona@imperial.ac.uk} \\
}

\begin{document}
\maketitle
\begin{abstract}
Converting process sketches into executable simulation models remains a major bottleneck in process systems engineering, requiring substantial manual effort, simulator-specific expertise, and iterative refinement. Although recent advances in generative AI have progressed both the automated interpretation of engineering diagrams and LLM-assisted flowsheet generation, these two lines remain largely disconnected: diagram-understanding methods typically stop at extracted graphs or semantic representations, while text-to-simulation workflows assume structured inputs rather than raw visual artifacts. Bridging this gap requires recovering structured process meaning from heterogeneous diagrams and instantiating simulator objects that satisfy strict creation, connectivity, and initialization rules. To address this, we present an end-to-end multi-agent large language model system that converts process diagrams directly into executable Aspen HYSYS flowsheets. The system decomposes the task across three coordinated layers: diagram parsing and interpretation, simulation model synthesis, and multi-level validation. Each layer consists of specialized agents handling visual interpretation, construction of a graph-based intermediate representation, code generation for the HYSYS COM interface, execution, and structural verification. This decomposition enables execution grounding and explicit error localization, reducing hallucination risk while keeping failure modes transparent. We evaluate the framework on four chemical engineering case studies of increasing complexity, from a simple desalting process to an industrial-scale aromatic production flowsheet with multiple recycle loops. The system produces executable HYSYS models in all cases, achieving complete structural fidelity (F1 = 1.00 across all metrics) on the two simpler cases and maintaining high performance on the more complex ones (connection consistency $\geq 0.93$, stream consistency $\geq 0.96$). Ablation analysis confirms that each architectural component contributes meaningfully to robustness, with sensitivity increasing alongside diagram complexity. These results establish a viable end-to-end sketch-to-simulation automation in process systems engineering, while indicating that remaining challenges lie primarily in handling dense recycle structures, implicit diagram semantics, and simulator-interface constraints.
\end{abstract}





\section{Introduction}\label{sec:1}
\subsection{Problem Context}
Process simulation stands as the computational backbone of modern chemical engineering practice, providing a sound basis for process design, analysis, and operational decision-making. Looking ahead, its importance is likely to grow further as emerging paradigms such as digital twins, real-time optimization, and more autonomous modes of operation depend increasingly on continuously updated, high-quality, reliable process models \cite{peterson_digital_2025}. 

However, developing a high-fidelity simulation model remains a major practical bottleneck. The process often requires time-intensive manual effort, simulator-specific expertise, and repeated refinement, with even minor structural or specification errors capable of preventing valid execution. These challenges are further amplified when the starting point is a high-level process diagram rather than a structured digital representation. Methods and tools that automate simulation model generation are therefore becoming increasingly important for enabling faster, more reliable engineering workflows and for supporting the broader shift towards digitalization and automation in process systems engineering (PSE) \cite{liang_large_2026, tian_text_2026}. 

\subsection{Research Gap}
Process diagrams encode valuable engineering information, including major unit operations, material streams, and the overall topology of a process system. As such, they serve as a standard high-level representation for communicating process structure and design intent. However, despite the value of this information, converting diagrams into executable simulation models remains a largely manual task: substantial interpretation, inference, and iterative refinement are required to yield a valid simulation \cite{towler_chemical_2013}.

A fundamental challenge arises from the gap between the information provided by process diagrams and the requirements of process simulation software. Diagrams are intended to support human understanding of process structure, whereas simulation environments require explicit, machine-interpretable definitions of units, stream connections, specifications, and initialization conditions. As a result, important information may be missing, implicit, or ambiguous in the source diagram. Therefore, converting such diagrams into executable models requires more than extraction; it requires engineering interpretation and a structured synthesis process that organizes, supplements, and translates diagram content into simulator-compatible form. 

This work is motivated by the need to bridge the gap between diagram understanding and automated model generation. Recent advances in computer vision, including optical character recognition and multimodal artificial intelligence, have improved the identification and extraction of symbols, labels, and visual relationships from technical diagrams \cite{bray_decoding_2026, shteriyanov_enhancing_2025}. In parallel, prior approaches to automated model generation have shown how structured inputs can be converted into executable models \cite{liang_large_2026, tian_text_2026}. However, these two lines of work remain only partially connected. The former generally ends at recognition and extraction, whereas the latter typically assumes explicit, structured process inputs rather than raw diagrams. As a result, end-to-end frameworks capable of transforming raw visual inputs into validated, executable process simulation models remain limited. 

\subsection{Objective, Scope, and Contributions}
To address the challenges of manual model development, this work proposes a multi-agent system for transforming process diagrams into executable process simulation models. The proposed workflow is designed to bridge the gap between high-level engineering diagrams and simulator-ready process models through coordinated stages of diagram interpretation, structured model generation, and validation. The scope of the study is restricted to steady-state chemical process simulation, with Aspen HYSYS used as the target environment for model construction and execution. 

Within this scope, the main contributions of this work are threefold. First, it presents an end-to-end multi-agent architecture for automated process model generation from visual diagrams in chemical engineering. Second, it introduces a structured intermediate representation for converting diagram-level information into simulator-compatible model elements. Third, it integrates automated model generation with validation and execution in Aspen HYSYS, and demonstrates the resulting workflow across four case studies of increasing complexity. 
 
The remainder of this paper is organized as follows: Section \ref{sec:2} reviews the background and related work on automated model generation. Section \ref{sec:3} presents the proposed multi-agent methodology. Section \ref{sec:4} describes the case studies used for evaluation. Section \ref{sec:5} presents the results and discusses workflow performance, robustness, and limitations. Finally, Section \ref{sec:6} concludes the work and outlines directions for future work. 

\section{Background and Related Work}\label{sec:2}
\subsection{Process Simulation Environments}
Process simulation environments are software platforms used to construct, specify, and execute computational models of process systems. Originating from early computer-aided flowsheeting tools, they have evolved into mature systems with integrated thermodynamic methods, unit-operation libraries, and graphical interfaces. In chemical engineering, they are widely used across process design, analysis, optimization, and operational support, making them a central part of modern engineering workflows \cite{towler_chemical_2013}. As computing capabilities advance, their role has expanded beyond steady-state design calculations to include dynamic simulation and broader digital engineering applications. 
 
In practice, process simulation platforms are often categorized by their underlying computational architecture. Commercial environments such as Aspen HYSYS, Aspen Plus, and AVEVA PRO/II typically follow a Sequential Modular (SM) approach, in which unit operations are solved individually in a specified sequence \cite{aspentech_aspen_2026, aveva_aveva_2026}. This architecture aligns closely with the visual logic of engineering diagrams. By contrast, platforms such as gPROMS employ an Equation-Oriented (EO) formulation, in which the full system of algebraic and differential equations is solved simultaneously \cite{siemens_gproms_2026}. Although EO platforms offer greater flexibility for high-fidelity modeling and complex optimization, SM environments remain the industrial standard for general process design and analysis because of their robustness and intuitive flowsheet-based construction \cite{dimian_integrated_2014}. 
 
Within this class of sequential-modular platforms, different environments have developed distinct areas of industrial emphasis. Aspen HYSYS is widely used and often preferred, particularly in oil and gas and broader energy applications, whereas Aspen Plus is more general-purpose and commonly associated with chemical process applications, including specialty chemicals and pharmaceuticals \cite{chukwu_optimising_2025}. AVEVA PRO/II occupies a similar steady-state simulation space and has long been used across refining and chemical applications. These distinctions are not absolute, but they illustrate how commercial simulation platforms with similar underlying architectures often evolve toward different sectoral strengths. 
 
A key differentiating feature of these environments is their support for external integration and automation. Modern commercial platforms expose their internal object models through application programming interfaces or interoperability layers such as COM (Component Object Model) or Python-based wrappers. These interfaces allow external software, including the agents considered in this research, to programmatically construct and modify flowsheets without manual graphical intervention \cite{kumar_integrating_2025, santos_bartolome_comparative_2022}. In addition, international standards such as CAPE-OPEN have sought to formalize interoperability across simulation platforms, enabling thermodynamic methods and unit-operation models to be exchanged more consistently between environments \cite{co-lan_cape-open_2026}. 
 
Within this landscape, Aspen HYSYS is selected as the target simulation environment for the present study. This choice is motivated by its extensive industrial use and strong exposure in practice, particularly within the broader energy and manufacturing industry, where it is widely used and often preferred for steady-state hydrocarbon process modeling. Although integrating automated workflows with a commercial platform such as HYSYS presents practical implementation challenges, including the orchestration of its COM-based object model, its industrial relevance ensures that the proposed framework is evaluated against a rigorous real-world benchmark. Consequently, HYSYS provides a meaningful testbed for assessing the feasibility of translating diagrammatic intent into executable, simulator-compatible flowsheets. 

\subsection{Prior Work on Diagram Understanding (Automated Interpretation of Engineering Diagrams)}
Prior work on engineering diagram understanding has evolved from low-level visual extraction toward increasingly structured and semantically informed forms of analysis. In chemical engineering, this progression has been driven by a transition from rule-based and classical vision techniques to deep-learning, graph-aware, and multimodal approaches that enable richer reasoning over process diagrams. 
 
The first stage focuses on element extraction. Early approaches treated process diagrams as visual documents whose basic components, such as text labels, symbols, arrows, and line objects, had to be identified separately. To achieve this, they typically combined OCR, template matching, Hough-transform-based line extraction, and rule-based visual processing to identify the visible building blocks of the diagram \cite{kang_digitization_2019, moon_deep_2021, rahul_automatic_2019}. While these methods performed well on relatively clean and standardized drawings, they were often sensitive to clutter, inconsistent notation, overlapping lines, and the variability common in legacy scans and industrial diagrams. More recent deep-learning-based approaches improved the robustness of this stage by learning symbol, text, and feature representations directly from data, often using convolutional neural networks or transformer-based vision models. This has been especially useful for dense or heterogeneous diagrams. Examples include deep-learning-based symbol and text recognition in high-density P\&IDs \cite{kim_deep-learning-based_2021}, feature recognition from image-format P\&IDs \cite{yu_features_2019}, and broader deep neural-network-based recognition of image-format P\&IDs \cite{su_image_2024}. 
 
The second stage extends beyond isolated recognition toward topology reconstruction. At this point, the objective is no longer only to detect visible elements, but to recover how they are connected and what process structure they represent. This includes associating labels with symbols, tracing pipelines, identifying stream connectivity, and representing the result as a graph or other digital process structure. End-to-end digitization frameworks such as \cite{kim_end--end_2022} and Digitize-PID \cite{paliwal_digitize-pid_2021} illustrate this transition by combining object recognition with topology reconstruction. Similar ideas appear in chemical PFD digitization, where deep-learning-based detection is followed by connectivity recovery to obtain a structured process representation \cite{theisen_digitization_2023}. More recent work also moves toward joint structural prediction and highlights the need to identify omissions and inconsistencies after the initial parsing stage \cite{kim_automated_2025}, reflecting a broader recognition that reliable engineering diagram interpretation often requires explicit validation when the output is intended for downstream technical use. 
 
The third stage focuses on semantic interpretation. Rather than stopping at the extracted structure, this stage treats the diagram as a semantic object that can support querying, higher-level reasoning, and downstream engineering use. This is reflected in recent multimodal and graph-grounded systems that transform static diagrams into structured knowledge representations and enable question answering or language-based interaction over them \cite{gupta_pidqaquestion_2025}. These methods mark a clear shift from perception and topology recovery toward semantic interpretation. However, even at this level, the output typically remains a graph, knowledge base, or semantic representation rather than a simulator-ready model specification. In particular, current systems may identify that a unit, stream, or connection exists without supplying the domain-specific parameters, initialization conditions, and simulator constraints required for executable model synthesis. 
 
Taken together, these stages reveal a clear pattern in the literature: the field has progressed from extracting visible elements, to reconstructing process topology, and more recently to interpreting engineering meaning. This progression is highly relevant to the present work. Diagram understanding is a necessary foundation for automated model generation, but it is not sufficient on its own. A significant gap remains between diagram interpretation and executable simulation models. The scope of current frameworks generally stops at extracted structures, graphs, or semantic representations, whereas simulation-model synthesis requires parameter inference, initialization logic, and simulator-specific mapping. In this sense, the remaining challenge is not only diagram understanding, but also bridging the symbol-to-parameter gap between recognized engineering elements and executable process-model definitions. 

\subsection{Prior Work on Automated Model Generation}

\subsubsection{Optimization and Reinforcement Learning Approaches}
Early work in this area emerged from classical process synthesis, where automated flowsheet generation was formulated as an optimization problem over a predefined superstructure of candidate units and connections. Within this framework, mathematical programming was used to select a feasible or optimal substructure from that design space 
\cite{grossmann_mixed-integer_1985,pistikopoulos_advanced_2024, westerberg_synthesis_1989}. Similarly, more recent symbolic approaches replace explicit superstructures with machine-readable encodings. For example, eSFILES represents process structures through symbolic flowsheet strings and uses these encodings as the basis for intelligent synthesis \cite{mann_esfiles_2024}. 

Rather than solving a fixed optimization problem directly, a second line of work casts flowsheet generation as a sequential decision problem. A reinforcement learning agent constructs the flowsheet step by step by adding units and connections within a predefined synthesis environment \cite{gottl_automated_2022}. This perspective has since been extended to include downstream design and control decisions, allowing generation to be coupled more closely with process performance and operability \cite{reynoso-donzelli_reinforcement_2025}. Although these methods enable more flexible exploration of the design space than classical optimization-based synthesis, they still rely on predefined actions, candidate units, and constraints. Moreover, reinforcement learning is primarily designed for learning decision policies through repeated interaction, rather than for one-off design synthesis tasks. In both cases, the process must first be represented in machine-readable form, and the output is typically a flowsheet configuration rather than a fully executable simulation model.

\subsubsection{Semi-Automatic, LLM, and Multi-Agent Approaches}
A transition toward semi-automatic modules can be seen in the work of \cite{sierla_towards_2020}, where a digitalized P\&ID is transformed into a directed graph and then into a simulator-specific flowsheet skeleton. However, the workflow remained semi-automatic, since experts still had to choose the appropriate simulator blocks for the generated structure and complete the initialization and parameter settings needed to finalize the model. 
 
LLMs and agentic systems extend automation further by shifting the interface from structured engineering data toward natural-language instructions and higher-level task orchestration. In the work \cite{liang_large_2026}, for example, an LLM agent is integrated with AVEVA Process Simulation to support natural-language interaction, guided flowsheet construction, data extraction, and optimization support. Their results show that step-by-step interaction is more reliable for guided model building, whereas single-prompt generation is faster but still requires expert oversight because of oversimplification and calculation errors. A more automated approach is proposed by \cite{tian_text_2026}, who formulated text-to-simulation as a multi-agent workflow with specialized agents for task understanding, topology generation, parameter configuration, and evaluation analysis. Their system reports improved convergence and reduced design time, but it still begins from textual process specifications as opposed to visual engineering inputs. 
 
A further extension appears in \cite{srinivas_autochemschematic_2025}, which adopts a more integrated, physics-aware framework for the scale-up of chemical manufacturing. Rather than acting only as a simulator assistant or text-to-simulation pipeline, it generates PFDs and P\&IDs from textual or retrieved process descriptions and validates them through a simulator-supported closed-loop workflow using DWSIM. This makes it one of the closest examples of closed-loop, agentic engineering generation, though it still starts from text or retrieved knowledge rather than from raw visual engineering diagrams. 
 
A related extension of this trend appears in CeProAgents, which proposes a broader multi-agent framework that spans knowledge retrieval, concept-level diagram reasoning, and simulator-based parameter optimization across the process development lifecycle. Conceptually, it is more integrated than simulator-assistant or text-to-simulation workflows, since it attempts to connect natural-language objectives, process-diagram abstractions, and Aspen-based optimization within one architecture. However, the released implementation appears closer to a collection of partially connected modules than to a fully seamless end-to-end pipeline, with stronger support for parsing and optimization than for complete abstract-to-simulator model generation. As a result, it is best understood as an ambitious step toward integrated agentic process development rather than a complete solution for converting raw engineering diagrams directly into executable simulation models \cite{yang_ceproagents_2026}. 
 
Collectively, this literature shows that automated model generation in chemical engineering is already well developed once the process has been expressed in a machine-readable form. Classical synthesis methods, symbolic representations, sequential generation frameworks, digital-twin pipelines, and agentic workflows all demonstrate different ways of automating flowsheet or model construction. The main limitation, however, is that most of these approaches begin only after the representation problem has already been resolved. They automate the transition from formal process description to flowsheet or simulator model, but not the earlier transition from ambiguous visual engineering input to a structured, simulator-ready representation. This is the distinction addressed in the present work, which connects visual interpretation with structured synthesis and validation within a single workflow.

\subsection{Multi-Agent Systems for Complex Engineering Workflows}
A multi-agent system (MAS) is a system composed of multiple autonomous agents that perceive their environment, make local decisions, and act in pursuit of individual or collective goals. An agent, in this context, is a computational entity capable of sensing relevant inputs, reasoning over them, and performing actions directed toward a defined objective.  The broader contemporary literature often discusses such goal-directed autonomous behavior under the term Agentic AI, particularly when agents operate with planning, tool use, and limited human supervision \cite{rupprecht_multi-agent_2026}.   

The defining feature of MAS is that problem solving is distributed rather than centralized: specialized agents interact through coordination, cooperation, or negotiation, and their collective behavior can achieve outcomes beyond those of isolated components. This makes MAS well-suited to complex, multi-step engineering problems that require distributed reasoning, modular task allocation, and coordination across various computational tools. 

In practice, MAS can take different forms depending on how knowledge, decision-making, and coordination are organized. Information may be distributed across agents or shared through a common state, with agents updating their local or global view of the problem as new information becomes available. Agents can be instances of rule-based logic, optimization routines, reinforcement-learning policies, or large language models, depending on the role they are intended to perform. Coordination may be imposed through a centralized orchestrator or arise from direct interaction among agents. As a result, MAS can support both deterministic workflows, where execution order is predefined, and more adaptive settings, where routing, sequencing, and task assignment evolve in response to intermediate results or changing system conditions. 

A key advantage of multi-agent systems lies in their ability to support specialization and coordination at the same time. Here, architecture refers to the organizational structure through which agents are arranged, responsibilities are allocated, and interactions are coordinated. Such architectures may be sequential, hierarchical, or orchestrated through a central controller, depending on how task dependencies and information flow are managed. Different agents can be assigned different reasoning modes, representations, or external tools, while higher-level coordination mechanisms allow their outputs to be combined into a coherent workflow. This is especially useful when solving problems that require both local task expertise and global consistency. Rather than forcing all subtasks into a single representation or model, a multi-agent architecture allows perception, structure recovery, parameter inference, and validation to be treated as related but distinct forms of reasoning. 
In chemical engineering, multi-agent systems have evolved across several generations. Early studies used agents as modular carriers of engineering knowledge, decomposing complex tasks into specialized decision units for process design and fault diagnosis \cite{eo_cooperative_2000, han_agent-based_1995}. Later work extended the scope toward interoperability, enterprise coordination, distributed optimization, refinery applications, and fault diagnosis in transient operations to solve more complex engineering tasks 
\cite{julka_agent-based_2002, seng_multi-agent_nodate, siirola_toward_2003, stalker_cogents_2004, yang_multi-agent_2008}.

More recent research has introduced learning-enabled architectures, particularly reinforcement-learning-based multi-agent systems, in which agents adapt their behavior in response to intermediate results and support data-driven coordination in tasks such as process control and scheduling 
\cite{hong_naphtha_2024, yue_multi-agent_2023}. Building on this shift toward adaptive computation, the latest contributions have expanded MAS further through agentic architectures that incorporate multimodal large language models, retrieval-augmented generation, graph-based retrieval, and tool use for tasks such as process improvement, industrial control, operational assistance, process optimization, and PFD/P\&ID generation from text or multimodal inputs 
\cite{du_potential_2025, gowaikar_agentic_2024, lee_gpt_2024, srinivas_autochemschematic_2025, tian_text_2026, vyas_autonomous_2025, zeng_llm-guided_2025}. 

This evolution is directly relevant to the present work. Transforming a visual engineering diagram into an executable simulation model is not a single inference task, but a sequence of heterogeneous subtasks that includes visual interpretation, structural reconstruction, simulator-specific mapping, specification completion, and execution-oriented validation. These subtasks differ not only in their inputs and outputs, but also in the type of reasoning they require. A multi-agent system, therefore, provides a natural architectural basis for this problem, because it allows these stages to be handled by specialized agents while still maintaining coordination across the overall workflow. 

For this reason, the present work adopts a multi-agent system not simply as an implementation choice, but as a methodological response to the structure of the problem itself. The proposed workflow uses specialized agents to bridge visual engineering inputs and simulator-ready model generation through coordinated interpretation, synthesis, and validation. 

\subsection{Research Gap and Positioning}
The literature has advanced along two directions that remain insufficiently integrated. On one side, engineering-diagram understanding has progressed from element extraction to topology reconstruction and semantic interpretation, but most methods still terminate at a graph or descriptive representation rather than an executable simulation model. On the other hand, recent multi-agent and agentic workflows have enabled increasingly capable text-to-simulation and simulator-assisted design, but they usually begin from textual specifications, structured engineering data, or standardized intermediate forms. The unresolved gap lies in connecting these directions: transforming raw visual engineering diagrams directly into simulator-ready models. 

A further limitation is that many existing workflows are tied either to specific input formats or to non-industrial modeling environments. In practice, however, engineering diagrams vary widely in notation, layout, completeness, and degree of standardization. A practically useful framework should therefore not rely exclusively on highly standardized exchange formats, but should be able to operate across a broader range of engineering diagrams, including those encountered in less structured industrial settings. At the same time, industrial relevance depends on compatibility with widely used commercial simulators rather than only research-oriented platforms or custom prototypes. For this reason, the present work targets visual engineering inputs more broadly and uses Aspen HYSYS as the simulation environment. 
 
This work is positioned at the intersection of these gaps. It proposes an end-to-end multi-agent workflow that begins from visual engineering diagrams rather than textual specifications and aims to generate executable simulation models rather than only topological or semantic representations. By combining diagram interpretation, structured synthesis, specification completion, and simulator-based validation within a single coordinated framework, the proposed system treats the simulator not as a passive output target but as an active component of the generation and verification loop. In this way, the framework seeks to connect visual engineering inputs to simulator-ready model generation in a form that is both technically executable and industrially relevant. 

\section{Methodology}\label{sec:3}
This work proposes a multi-agent system for transforming process diagrams into executable process simulation models. This section presents the methodology underlying the proposed framework, beginning with the overall system architecture, followed by detailed descriptions of the interpretation, synthesis, and validation layers. It then outlines the rationale for model selection and deployment, and concludes with the system implementation. 

\subsection{Multi-Agent System Architecture}
The automated transformation of process diagrams into ready-to-use process simulation models is a complex reasoning task; it requires visual interpretation, information extraction, semantic parsing, and model synthesis, each with distinct logic and failure modes. Monolithic approaches built around a single LLM agent, therefore, tend to lose robustness as diagram complexity increases. More concretely, in multi-step reasoning tasks, as intermediate states accumulate within a single context window, attention allocation and context utilization become less reliable over dispersed evidence, while composing multiple dependent subtasks within a single reasoning path makes errors harder to isolate, trace, and correct \cite{brinkmann_mechanistic_2024, dziri_faith_2023, liu_lost_2024}.

\begin{figure}[H]
    \centering
    \includegraphics[width=\linewidth]{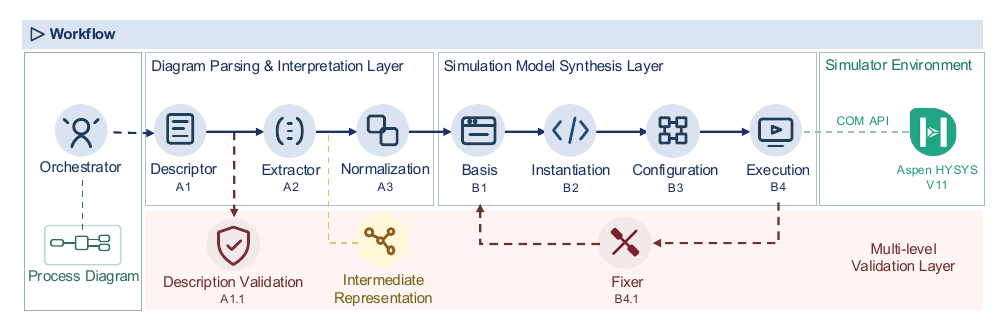}
    \caption{Multi agent system architecture}
    \label{fig:methodology}
\end{figure}

To address these limitations, this study proposes a modular multi-agent system architecture, motivated by prior literature that emphasizes the value of task decomposition, specialization, and error isolation in complex reasoning workflows \cite{guo_large_2024}. The architecture is organized into three functional layers, each composed of specialized agents. As shown in Figure \ref{fig:methodology}, the Diagram Parsing and Interpretation Layer transforms the visual input into a structured representation. The Simulation Model Synthesis Layer then translates this representation into an executable process model in the simulator environment. Throughout the workflow, the Multi-level Validation Layer applies targeted validation procedures to assess structural and semantic consistency at key stages. This layered design improves reliability, supports systematic error localization, and limits error propagation across stages.
Workflow execution is governed by a central orchestrator that manages state transitions between agents in a deterministic sequence. Each agent operates as a state transformer: it consumes a validated state from the preceding stage and returns a refined state for the next. This design enforces controlled and reproducible information flow while enabling targeted trace-based issue localization at any point in the workflow \cite{deshpande_trail_2025}.
Formally, let $s_k$ denote the workflow state after stage $k$, and let $A_k$ denote the transformation implemented by the agent at that stage. The workflow then evolves as
\[
s_{k+1} = A_k(s_k),
\]
where the orchestrator defines the ordered sequence of transformations $\{A_1, A_2, \ldots, A_n\}$. The final workflow state can therefore be written as
\[
s_n = (A_n \circ A_{n-1} \circ \cdots \circ A_1)(s_0).
\]

By assigning well-defined responsibilities to specialized agents, the architecture improves traceability, enables targeted validation, and supports modular extension. It thereby enables automated process model synthesis directly from process diagrams, while also allowing additional capabilities to be incorporated without redesign, supporting future extensions toward broader automated process design workflows.

\subsection{Diagram Parsing and Interpretation Layer}
The Diagram Parsing and Interpretation Layer is the entry point of the workflow. Process flow diagrams encode rich design information through symbols, annotations, stream labels, and spatial arrangements. However, they often exhibit visual clutter, inconsistent formatting, non-standard conventions, and implicit connectivity assumptions, which make direct automated interpretation challenging. The role of this layer is to transform the unstructured visual input into a formalized Intermediate Representation (IR) that captures the process topology in a structured, simulator-compatible form.
Formally, the IR is represented as a directed graph, $G=(V,E)$, where $V$ denotes the set of unit operations and $E\subseteq V\times V$ denotes the directed material streams connecting source and destination units. In implementation, this graph is instantiated as a structured JSON schema comprising two principal collections: a set of units representing graph nodes, $V$, and a set of material streams representing directed edges, $E$. This representation preserves flow directionality and provides a clear interface to downstream synthesis agents.

The interpretation layer workflow comprises three agents: the Descriptor Agent (A1), the Extractor Agent (A2), and the Normalization Agent (A3), each discussed in the following subsections.

\subsubsection{Descriptor Agent (A1)}
The Descriptor Agent (A1) uses a multimodal LLM to generate a detailed description of the process depicted in the input diagram. Prior to inference, the diagram is standardized to a fixed resolution to improve computational efficiency while preserving the spatial relationships required for process interpretation. The prompt enforces a left-to-right traversal heuristic and instructs the model to explicitly enumerate all visible process elements.

To improve consistency, the prompt incorporates a self-evaluation instruction that encourages the model to assess whether each described element is supported by observable visual evidence in the diagram 
\cite{madaan_self-refine_2023, weng_large_2023}. The resulting description captures the overall process intent, identified unit operations, material streams, and their inlet-outlet relationships.

\subsubsection{Extractor Agent (A2)}
Building on the output of Agent A1, the Extractor Agent (A2) employs a multimodal LLM to construct the JSON-based Intermediate Representation. Both the original diagram and description produced by Agent A1 are parsed by this agent. The diagram remains the primary source of ground truth, while the generated description serves as a semantic prior that helps reduce ambiguity and improve extraction reliability.

To improve robustness, extraction is performed in two inference passes. In the first pass, the agent identifies all visible unit operations and assigns unique identifiers. In the second pass, it extracts material streams, classifies them as feed, intermediate, or product streams, and restricts source-destination assignments to the unit set established in the first pass.

This staged extraction procedure strengthens topological consistency and mitigates common LLM failure modes, including equipment hallucination, label misinterpretation, and inconsistent connectivity assignments. It also localizes extraction errors, thereby limiting cross-stage error propagation. The output of Agent A2 is a structured Intermediate Representation that encodes the extracted process topology.

\subsubsection{Normalization Agent (A3)}\label{sec:3.2.3}
Before the IR is passed to the synthesis layer, it is processed by the Normalization Agent (A3), a rule-based agent that enforces topological consistency and simulator-specific structural requirements.

One responsibility of Agent A3 is to resolve implicit junctions commonly found in process diagrams. In many process diagrams, multiple streams converge directly into unit operations that are designed to accept a single inlet stream, such as pumps or compressors. Although visually intuitive, such representations violate the nodal constraints imposed by the simulator. Agent A3 detects these multi-stream convergences and inserts explicit mixing or splitting units into the IR, rerouting the associated streams to preserve process intent while satisfying simulator requirements.

A second responsibility of Agent A3 is to align extracted process structures with simulator templates. Certain unit operations may appear in the diagram as multiple functional elements, but must be instantiated in the simulator as a single object. Distillation systems are an example of this: the column tower, condenser, and reboiler may be extracted as separate elements, but they must be consolidated into a predefined unit operations template. Agent A3 therefore restructures the IR into a cohesive, template-compliant representation that can be instantiated as the corresponding simulator object.
At the conclusion of the interpretation layer, the system produces a normalized, simulator-compatible Intermediate Representation of the process flowsheet. This IR then serves as the controlled input to the synthesis layer, where it is translated into an executable process simulation model.

\subsection{Simulation Model Synthesis Layer}
The Simulation Model Synthesis Layer translates the formalized Intermediate Representation into an executable process simulation model in Aspen HYSYS. To achieve this, four specialized coding agents sequentially construct a Python automation script that interacts with the Aspen HYSYS Component Object Model (COM) interface. The agents operate on a shared Python template and modify only predefined execution blocks. This guided synthesis strategy enforces adherence to simulator-specific conventions.

The synthesis layer workflow comprises four agents: the Basis Agent (B1), the Instantiation Agent (B2), the Configuration Agent (B3), and the Execution Agent (B4), each discussed in the following subsections.

\subsubsection{Basis Agent (B1)}
The Basis Agent (B1) is a code-oriented LLM agent that establishes the simulation case basis by defining the case name, selecting the appropriate fluid property package, and constructing the component list required for the process. A key responsibility of the agent is translating the extracted feed components into valid Aspen HYSYS component names. This translation is performed using a Retrieval-Augmented Generation (RAG) module that queries a curated knowledge base to align extracted material names with exact entries in the HYSYS pure-component database, thereby ensuring that only simulator-compatible components are introduced into the simulation environment.

\subsubsection{Instantiation Agent (B2)}
Once the simulation basis is established, the Instantiation Agent (B2), a code-oriented LLM agent, constructs the structural skeleton of the process flowsheet. Operating directly on the IR graph, it translates the node set V and edge set $E$ into executable code.
Using the Aspen HYSYS COM automation interface, the agent maps unit operations and material streams to their corresponding simulator object classes. Dedicated instruction files guide object creation patterns for each supported unit type, ensuring compliance with the simulator object hierarchy and preventing unsupported operations. By the end of this stage, the structural flowsheet has been instantiated within the simulation environment.

\subsubsection{Configuration Agent (B3)}
Building on the instantiated flowsheet, the Configuration Agent (B3), also a code-oriented LLM agent, establishes the topological connectivity of the process model. Guided by the source-destination relationships encoded in the edge set E, it links material streams to the appropriate inlet and outlet ports of each unit operation through the COM automation interface. Unit-specific instruction files govern the connection logic for each object class, ensuring that streams are attached to valid ports in a simulator-consistent manner. As a result, the generated automation script defines a fully connected process model.

\subsubsection{Execution Agent (B4)}\label{sec:3.3.4}
The synthesis layer concludes with the Execution Agent (B4), a hybrid agent that combines a rule-based execution step with an LLM-based fixing step. The rule-based step executes the generated Python automation script without modifying the underlying model logic and records solver status, runtime diagnostics, and execution logs. If execution fails, a code-oriented LLM agent analyzes the execution trace and applies targeted corrections to the script before reattempting execution. 

This stage serves as an execution-level validation step, confirming whether the generated model can run successfully within the simulator environment while supporting systematic issues tracing of the automated modeling workflow. It therefore completes the transformation from a structured process representation into an executable process simulation model.

\subsection{Multi-Level Validation Layer}
To improve the reliability, robustness, and traceability of the proposed framework, validation is embedded at multiple levels of the workflow. These validation procedures do not replace the core generative steps; rather, they act as diagnostic checkpoints that identify inconsistencies, improve transparency, and support error localization. A summary of these procedures is provided in Table \ref{tab:validation}.

\begin{table}[H]
  \centering
  \caption{Summary of validation mechanisms used in the proposed workflow}
  \label{tab:validation}
  \begin{tabularx}{\textwidth}{XXX}
    \toprule
    Mechanism & Purpose & Stage \\
    \midrule
    Description validation (A1.1) & Assess alignment between the visual input and the generated description & Post-descriptor (A1) \\
    Schema and prompt safeguards & Enforce structured outputs, object constraints, and internal consistency & Interpretation and synthesis agents \\
    Execution validation and fixing & Validate executability and apply targeted runtime corrections & Execution (B4) \\
    \bottomrule
  \end{tabularx}
\end{table}

Within the interpretation layer, an auxiliary Description Validation Agent (A1.1) evaluates the output of the primary Descriptor Agent (A1). Operating on the original diagram, this secondary LLM acts as an independent evaluator that assesses whether the generated process description is aligned with the visual content of the diagram. Although it does not modify the description itself, it provides a confidence signal regarding its consistency with the source diagram.

Additional safeguards are embedded directly within the prompts and output constraints used across the workflow. In the interpretation layer, agents operate under strict JSON schemas to enforce properly structured intermediate outputs. In the synthesis layer, coding agents follow guided instruction files and constrained templates to maintain adherence to simulator requirements. Furthermore, internal consistency checks are embedded in selected prompts to encourage self-assessment before final output generation.

Finally, end-to-end validation is performed during model execution, as discussed in Section \ref{sec:3.3.4}. At this stage, execution logs, solver diagnostics, and correction outcomes provide a final check on model executability and support error localization. Together, these procedures strengthen the reliability and traceability of the overall workflow.

\subsection{Model Selection and Deployment}
Model selection follows three design criteria: (i) modality alignment, ensuring that each model matches the form of its input data; (ii) computational efficiency, prioritizing models that enable reliable and scalable inference; and (iii) data governance, requiring sensitive simulation logic to remain within a controlled local environment. Based on these criteria, the interpretation and synthesis layers are assigned to different model classes and deployment environments.

\subsubsection{Interpretation Models and Cloud Deployment}
The interpretation layer, comprising the Descriptor Agent (A1) and Extractor Agent (A2), uses Gemini 3 Flash, a proprietary multimodal LLM capable of joint reasoning over images and text \cite{google_deepmind_gemini_2025}. This capability is essential for interpreting process diagrams, which require scientific visual reasoning and strong optical character recognition (OCR) for the proper interpretation of symbols, stream labels, spatial connectivity, and textual annotations.

This model selection is further supported by reported benchmark performance on demanding scientific reasoning tasks. Gemini 3 Flash achieved $81.2\%$ on the MMMU-Pro benchmark and $90.4\%$ on GPQA Diamond, both of which evaluate reasoning over complex visual and scientific content 
\cite{google_deepmind_gemini_2025, rein_gpqa_2023, yue_mmmu-pro_2025}. These capabilities are directly relevant to the interpretation of process diagrams. In comparison, the open-weight multimodal alternatives considered in this study, such as Qwen 3.5 series, showed lower performance on these benchmark categories \cite{qwen_team_qwen35_2026}. The interpretation layer is deployed in a cloud environment since multimodal inference is computationally intensive. 

By contrast, the Normalization Agent (A3) operates locally and applies deterministic rule-based transformations to enforce structural consistency and simulator-specific requirements. Since this stage is algorithmic rather than generative, it does not require language model inference.

\subsubsection{Synthesis Models and Local Deployment}
To comply with industrial data-sensitivity requirements established by the project partner, the synthesis and execution tasks are deployed locally. This ensures that proprietary simulation structures and generated automation scripts remain within a controlled execution environment. With the exception of one confidential instruction file, the code used to reproduce the reported results is made available.

For the foundational code-synthesis tasks handled by the Basis Agent (B1) and Instantiation Agent (B2), the system uses Qwen2.5-Coder-7B, \cite{hui_qwen25-coder_2024}. This model was selected for its efficient local inference and reliable structured code generation, while also avoiding unnecessary computational cost.

For more demanding code reasoning tasks, the Configuration Agent (B3) and the fixing component associated with the Execution Agent (B4) use Qwen3-Coder-30B, \cite{yang_qwen3_2025}. These tasks require stronger multi-step reasoning to interpret connectivity relationships between unit operations, resolve simulator-specific dependencies, and analyze runtime errors during automated error tracing.

This hybrid deployment strategy isolates computationally intensive multimodal interpretation tasks in the cloud environment while ensuring that proprietary simulation synthesis remains secure within a local execution environment. Table \ref{tab:model_selection_workflow} summarizes the deployed models across the workflow.

\renewcommand{\arraystretch}{1.5}
\begin{table}[H]
\centering
\caption{Model selection and deployment across workflow components}
\label{tab:model_selection_workflow}
\begin{tabularx}{\textwidth}{
    >{\raggedright\arraybackslash}p{4.2cm}
    X
    >{\raggedright\arraybackslash}p{2.8cm}
    >{\centering\arraybackslash}p{1.8cm}
}
\toprule
Agent & Model / logic core & Agent type & Environment \\
\midrule
Descriptor (A1)             & Gemini 3 Flash                      & Multimodal LLM    & Cloud \\
Validation (A1.1)           & Gemini 3 Flash                      & Multimodal LLM    & Cloud \\
Extractor (A2)              & Gemini 3 Flash                      & Multimodal LLM    & Cloud \\
Normalization (A3)          & Rule-based logic                    & Rule-based agent  & Local \\
Basis (B1)                  & Qwen2.5-Coder-7B                    & Code-oriented LLM & Local \\
Instantiation (B2)          & Qwen2.5-Coder-7B                    & Code-oriented LLM & Local \\
Configuration (B3)          & Qwen3-Coder-30B                     & Code-oriented LLM & Local \\
Execution and Fixing (B4)   & Rule-based logic + Qwen3-Coder-30B & Hybrid            & Local \\
\bottomrule
\end{tabularx}
\end{table}
\renewcommand{\arraystretch}{1}

\subsection{System Implementation}
The multi-agent workflow is implemented in Python and orchestrated using LangGraph, which models the system architecture as a directed computational graph \cite{langchain_ai_langgraph_2024}. Within this structure, nodes represent individual workflow agents, including both LLM-based and rule-based agents, while edges define the execution sequence and data dependencies.

Intermediate data is encapsulated in structured state objects that propagate systematically between agents. These states contain the evolving workflow state, including extracted unit operations, material streams, normalized intermediate representations, and generated simulation code. Each agent reads the current state, applies its designated transformation, and returns an updated state to the subsequent node.

At the agent level, LangChain provides abstractions for prompt construction, message handling, and structured output parsing. The implementation deliberately separates probabilistic reasoning from algorithmic execution: LLMs are invoked only for tasks that require model-based reasoning, such as diagram interpretation and code synthesis, while rule-based agents operate directly on the workflow state through deterministic Python transformations.

Model inference is managed through Ollama, which provides a unified serving interface for both local and cloud-based models \cite{ollama_team_ollama_2023}. Locally hosted code-synthesis models are served through Ollama on dedicated hardware to satisfy industrial data-governance constraints, while the multimodal interpretation model is accessed through the Ollama cloud tier. Using a common serving interface simplifies integration within the orchestration layer and reduces infrastructure complexity.

Aspen HYSYS was selected as the simulation environment because it is widely regarded as the industry gold standard for steady-state process modeling and provides a programmable Component Object Model (COM) automation interface suitable for script-based flowsheet generation. Through this interface, the system can synthesize flowsheets, configure process connectivity, and execute simulations without manual interaction with the graphical user interface.

The system operates on the dedicated workstation environment summarized in Table \ref{tab:computational_environment}. Together, these software and hardware components operationalize the multi-agent framework described in Section 3.1, enabling automated diagram interpretation, model synthesis, validation, and simulation execution within a unified workflow.

\begin{table}[H]
\centering
\caption{Computational environment and software stack}
\label{tab:computational_environment}
\begin{tabularx}{\textwidth}{
    >{\raggedright\arraybackslash}p{3.2cm}
    >{\raggedright\arraybackslash}p{3.8cm}
    X
}
\toprule
Category & Component & Specification \\
\midrule
Hardware environment & CPU & Intel Xeon Gold 6442Y (2.60\,GHz) \\
 & Memory & 256\,GB RAM \\
 & GPU & NVIDIA RTX A4000 (16\,GB VRAM) \\
 & Operating system & Windows 11 (64-bit) \\
 & Compute platform & CUDA 12.8 \\
\midrule
Software stack & Programming language & Python 3.11.13 \\
 & Agent framework & LangChain \\
 & Agent orchestration & LangGraph 1.0.9 \\
 & LLM serving framework & Ollama 0.17.4 (local and cloud) \\
 & Process simulator & Aspen HYSYS V11 \\
\bottomrule
\end{tabularx}
\end{table}
\section{Case Studies}\label{sec:4}
The multi-agent system is evaluated using four case studies representing common chemical engineering processes. The selected diagrams span increasing levels of process and topological complexity, including variations in the number of unit operations, stream interconnectivity, labeling clarity, and layout density. The case studies, shown in Table \ref{tab:cs} are presented in order of increasing complexity to enable a systematic evaluation of the robustness, scalability, and structural reasoning capability of the workflow.

\begin{table}[htbp]
\centering
\caption{Characteristics of the case study diagrams}
\label{tab:cs}
\begin{tabularx}{\textwidth}{c
                                >{\raggedright\arraybackslash}p{3.0cm}
                                >{\centering\arraybackslash}p{2.2cm}
                                >{\centering\arraybackslash}p{1.8cm}
                                >{\centering\arraybackslash}p{1.6cm}
                                X}
\toprule
Case & Process & Unit operations & Stream density & Recycle loops & Diagram characteristics \\
\midrule
1 & Desalting                 & Low      & Sparse   & None     & Missing labels, implicit mixing \\
2 & Merox Sweetening          & Moderate & Moderate & One      & Compact layout, ambiguous connectivity \\
3 & Atmospheric Distillation  & Moderate & Moderate & One      & Non-standard symbols, partially labeled units \\
4 & Aromatic Production       & High     & Dense    & Multiple & Industrial-scale flowsheet with complex interconnections \\
\bottomrule
\end{tabularx}
\end{table}

Taken together, the four case studies provide a structured evaluation across progressively increasing levels of process and topological complexity. The selected diagrams range from a simple baseline process to industrial-scale flowsheets with dense interconnections and multiple recycle loops. This progression enables a systematic assessment of the system’s robustness, scalability, and limitations when applied to diverse process  diagrams encountered in chemical engineering practice.

\subsection{Case Study 1: Desalting Process}
\begin{figure}[H]
    \centering
    \includegraphics[width=0.8\linewidth]{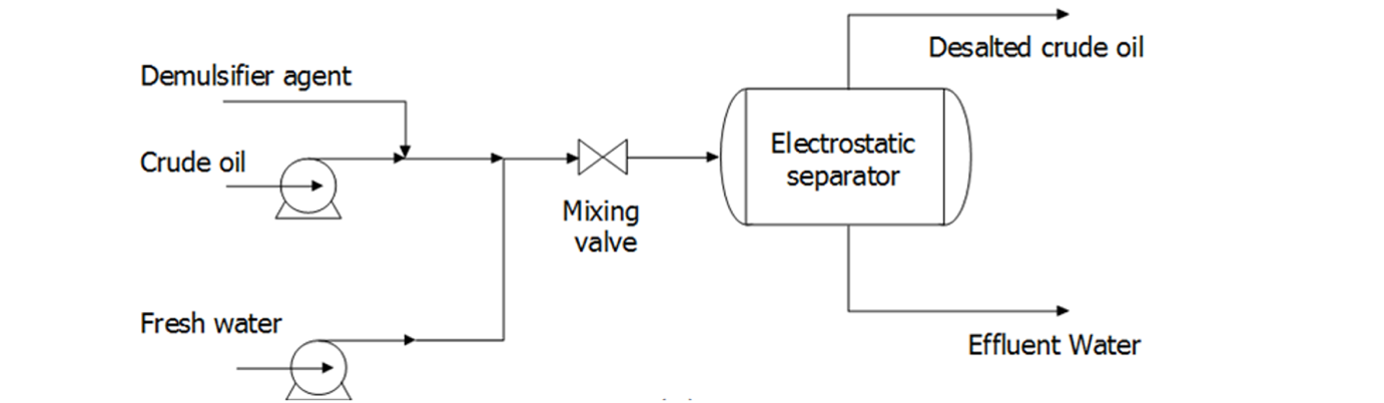}
    \caption{Desalting process (\cite{pereira_crude_2015})}
    \label{fig:cs1}
\end{figure}

The first case study considers a simplified crude oil desalting process obtained from a published process diagram (Figure \ref{fig:cs1}) \cite{pereira_crude_2015}. In this process, crude oil and fresh water are pressurized by dedicated pumps, combined with a demulsifier agent, and routed to an electrostatic separator that produces desalted crude oil and effluent water.

This case serves as a baseline scenario due to its low structural complexity and limited number of unit operations. Despite its simplicity, the diagram presents several interpretation challenges, including unlabeled pumps, an implicitly represented mixing operation, and missing stream labels. These features test the workflow’s ability to correctly identify equipment and infer stream connectivity under minimal topological complexity.

\subsection{Case Study 2: Jet Fuel Sweetening (Merox) Process}

\begin{figure}[H]
    \centering
    \includegraphics[width=0.8\linewidth]{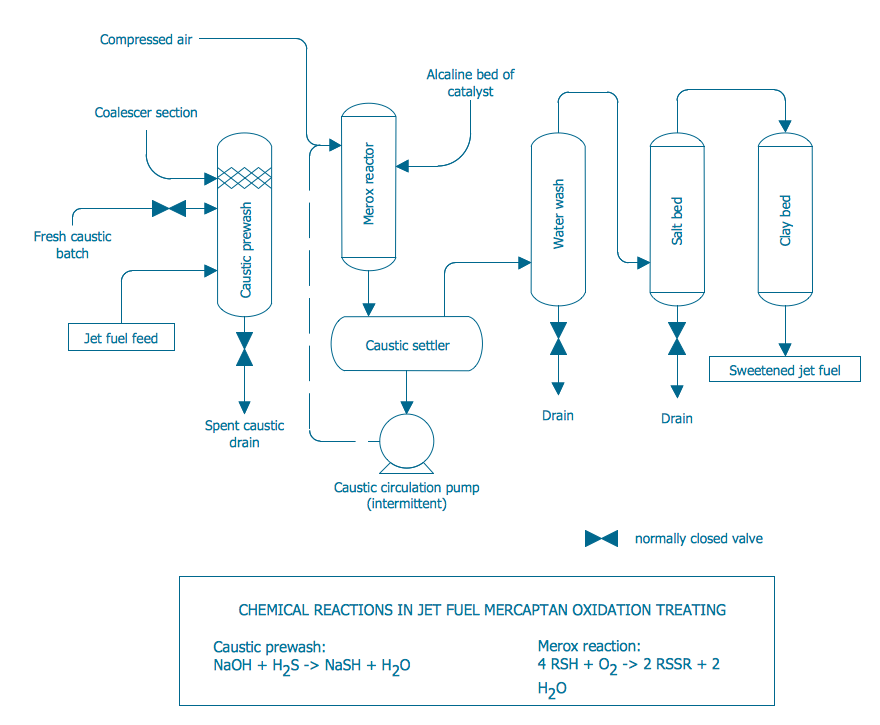}
    \caption{Merox process (\cite{conceptdraw_jet_nodate})}
    \label{fig:cs2}
\end{figure}

The second case study examines a jet fuel mercaptan oxidation treating process, commonly referred to as the Merox process. The corresponding process flow diagram was obtained from a publicly available source drawn using ConceptDraw, a diagramming platform (Figure \ref{fig:cs2}) \cite{conceptdraw_jet_nodate}.

The feed enters a caustic prewash vessel, after which it is routed to the Merox reactor, where mercaptan oxidation occurs in the presence of an alkaline catalyst and compressed air. Reactor effluent flows to a caustic settler for phase separation, after which the hydrocarbon stream passes through water washing, salt bed drying, and clay bed polishing units before exiting as the final product. A portion of the aqueous caustic phase is recycled to maintain caustic strength.

Relative to the baseline case, this flowsheet introduces moderate topological complexity. Although it follows standardized conventions, the flowsheet includes an internal recycle loop and ambiguous stream connectivity. In addition, dense textual annotations and reaction equations increase visual clutter, requiring the workflow to distinguish structural elements from explanatory content. This combination of features provides a useful stress test for the framework’s ability to interpret compact layouts and infer non-linear flow paths while maintaining structural consistency.

\subsection{Case Study 3: Atmospheric Crude Oil Distillation Process}

\begin{figure}[H]
    \centering
    \includegraphics[width=0.75\linewidth]{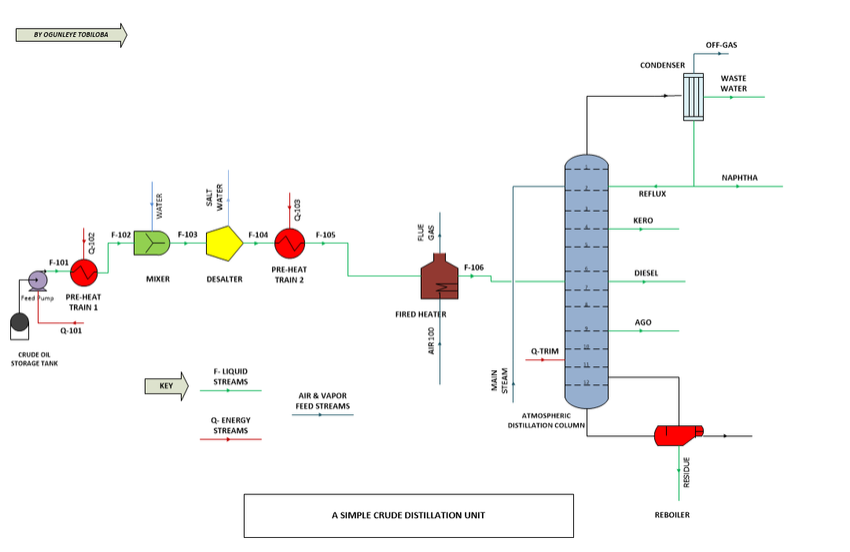}
    \caption{Crude distillation process (\cite{ogunleye_modeling_2021})}
    \label{fig:cs3}
\end{figure}

The third case study considers a classical atmospheric crude oil distillation process sourced from an undergraduate chemical engineering thesis (Figure \ref{fig:cs3})
\cite{ogunleye_modeling_2021}. The diagram represents a crude oil processing sequence consisting of crude preheating, desalting, fired heating, and atmospheric distillation.

In this process, crude oil is withdrawn from a storage tank and pressurized by a feed pump before entering a preheating train. The preheated crude is mixed with wash water and routed to a desalter, where salts and entrained water are removed. The desalted crude then passes through a second preheating train and a fired heater before entering the atmospheric distillation column. Within the column, the feed is separated into multiple fractions, including overhead products, side draws such as naphtha, kerosene, diesel, and atmospheric gas oil, and a bottom residue stream.

Compared with the previous cases, this example introduces additional interpretation challenges due to its non-standard formatting and limited equipment labeling. While process streams are identified, most equipment items are not explicitly labeled, creating ambiguity in unit recognition. The diagram also incorporates color-coded elements and unconventional symbols that deviate from standardized industrial flowsheet conventions. As a result, the workflow must rely more heavily on spatial relationships and contextual cues to infer unit roles and connectivity, thereby highlighting sensitivity to diagram quality rather than process complexity alone.

\subsection{Case Study 4: Aromatic Production Process}
\begin{figure}[H]
    \centering
    \includegraphics[width=0.8\linewidth]{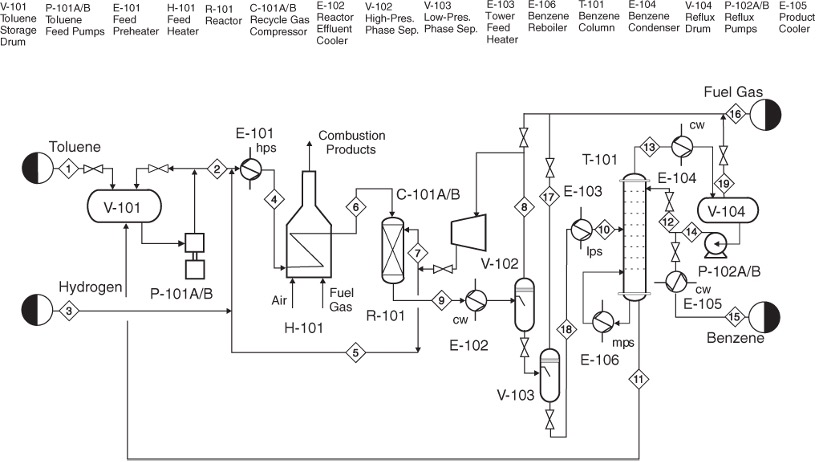}
    \caption{Aromatic production process (\cite{turton_analysis_2009})}
    \label{fig:cs4}
\end{figure}

The final case study considers a fully integrated, industrial-scale aromatic production process, obtained from a chemical engineering design textbook (Figure \ref{fig:cs4}) \cite{turton_analysis_2009}.The flowsheet encompasses a dense network of reactor systems, separation columns, heat exchangers, rotating equipment, and multiple recycle streams.

In this process, toluene is withdrawn from a storage drum and pressurized by feed pumps before being heated in a feed preheater and feed heater. The heated feed enters a reactor, where the primary reaction occurs in the presence of hydrogen. Reactor effluent is cooled and separated in a high-pressure separator, with part of the vapor phase compressed and recycled to the reactor. The liquid stream flows to a low-pressure separator, after which the hydrocarbon stream is routed to a benzene distillation column. Overhead vapor from the column is condensed and collected in a reflux drum, where a portion is returned as reflux while the remainder is withdrawn as benzene product. A reboiler provides heat input to maintain column separation.

Among the selected case studies, this flowsheet exhibits the highest level of process and topological complexity. The flowsheet contains many unit operations, dense stream interconnections, and multiple recycle loops linking reaction and separation sections. The compact arrangement of equipment and numerous crossing streams increases visual layout density and makes connectivity more difficult to interpret. Accordingly, this case tests the framework’s ability to maintain global structural consistency across large, interconnected flowsheets, thereby providing a realistic representation of industrial process diagrams.

\section{Results and Discussion}\label{sec:5}
This section presents the results of applying the multi-agent system across four case studies. It first defines the evaluation criteria, then discusses overall workflow performance and ablation results. It next examines model behavior, robustness, and variability at the multimodal reasoning level, and concludes with practical limitations and a summary of key findings.

\subsection{Evaluation Criteria}\label{sec:5.1}
A combination of quantitative and qualitative criteria is used to systematically assess the performance of the multi-agent system. The quantitative metrics measure structural fidelity relative to reference diagrams, while the qualitative criteria assess model behavior, robustness, and failure characteristics observed across the four case studies.
For the quantitative analysis, four metrics are defined to evaluate the accuracy of a process simulation model: Unit Consistency (UC), Stream Consistency (SC), Connection Consistency (CC), and Material Consistency (MC), as summarized in Table \ref{tab:quantitative_metrics}.

\begin{table}[htbp]
\centering
\caption{Quantitative evaluation metrics}
\label{tab:quantitative_metrics}
\begin{tabularx}{\textwidth}{>{\raggedright\arraybackslash}p{3.2cm}
                                >{\raggedright\arraybackslash}p{4.2cm}
                                X}
\toprule
Metric & Evaluated element & Purpose \\
\midrule
Unit Consistency (UC)       & Unit operations                   & Evaluates the correctness of extracted unit operations \\
Stream Consistency (SC)     & Material streams                  & Evaluates the correctness of extracted material flows \\
Connection Consistency (CC) & Directed unit-to-unit connectivity & Evaluates the correctness of process topology \\
Material Consistency (MC)   & Feed and process material components & Evaluates the correctness of extracted material and component identities \\
\bottomrule
\end{tabularx}
\end{table}

Each consistency metric is computed using the F1-score formulation \cite{van_rijsbergen_information_1979}:

$$
F1 = \frac{2PR}{P + R}  
$$

\pagebreak
The F1-score provides a harmonic mean of precision and recall, ensuring balanced penalization of both error types. For a given structural element set $X$ (units, streams, connections, or materials), precision ($P$) represents the proportion of extracted elements that are correct, and recall ($R$) represents the proportion of reference elements that are successfully extracted, defined as:

$$
P = \frac{TP}{TP + FP}, \quad
R = \frac{TP}{TP + FN}
$$

where True Positives, $TP$, denote correctly extracted elements present in the reference diagram, False Positives, $FP$, denote extracted elements not present in the reference diagram (hallucinated elements), and False Negatives, $FN$, denote reference elements that were not extracted (missing elements). This formulation ensures that omissions and hallucinations are penalized simultaneously, rather than rewarding structural completeness alone.
The quantitative metrics alone do not fully capture system behavior; therefore, qualitative analysis is also used to interpret the results. Section \ref{sec:5.2} evaluates performance across the four case studies, focusing on structural accuracy and execution stability. Section \ref{sec:5.3} then examines the contribution of individual agents through controlled ablation analysis. Section \ref{sec:5.4} analyzes model behavior, reproducibility, and multimodal benchmarking. Finally, Section \ref{sec:5.5} discusses practical limitations and deployment considerations. Together, these analyses provide a comprehensive assessment of the system.

\subsection{Overall Performance}\label{sec:5.2}
\begin{figure}[H]
    \centering
    \includegraphics[width=0.9\linewidth]{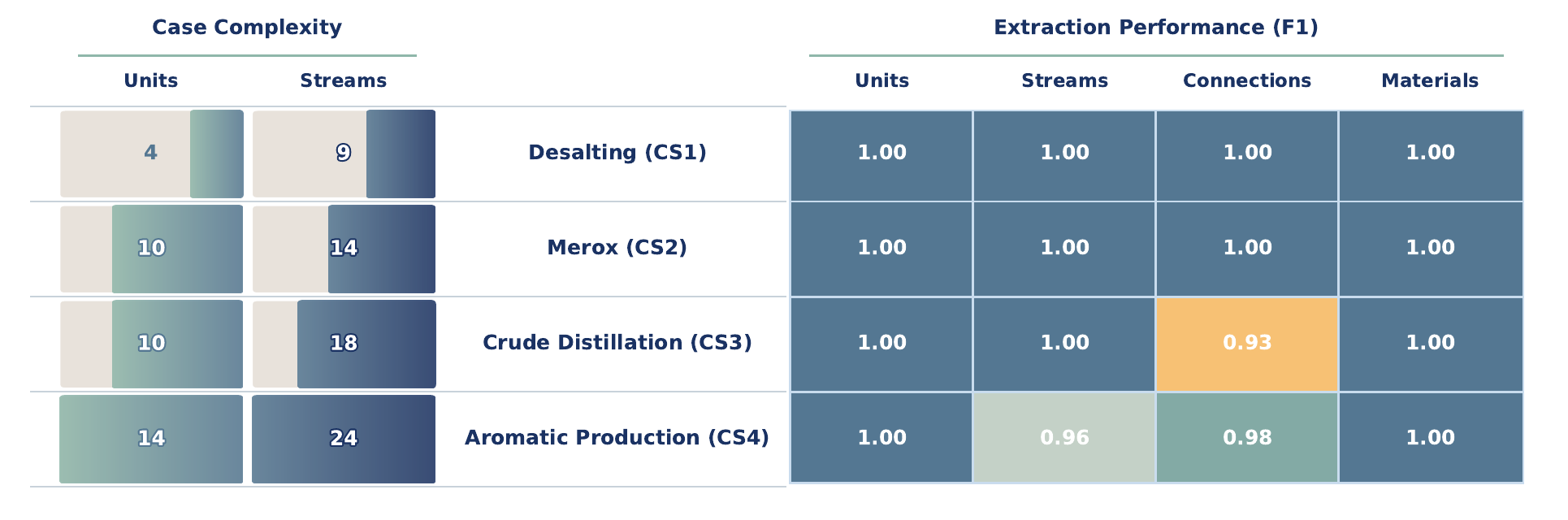}
    \caption{Overall structural performance across the four case studies}
    \label{fig:ovr}
\end{figure}

Across all four case studies, the multi-agent system successfully generated executable process simulation models. As summarized in Figure \ref{fig:ovr}. 

\begin{figure}[H]
    \centering
    \includegraphics[width=0.8\linewidth]{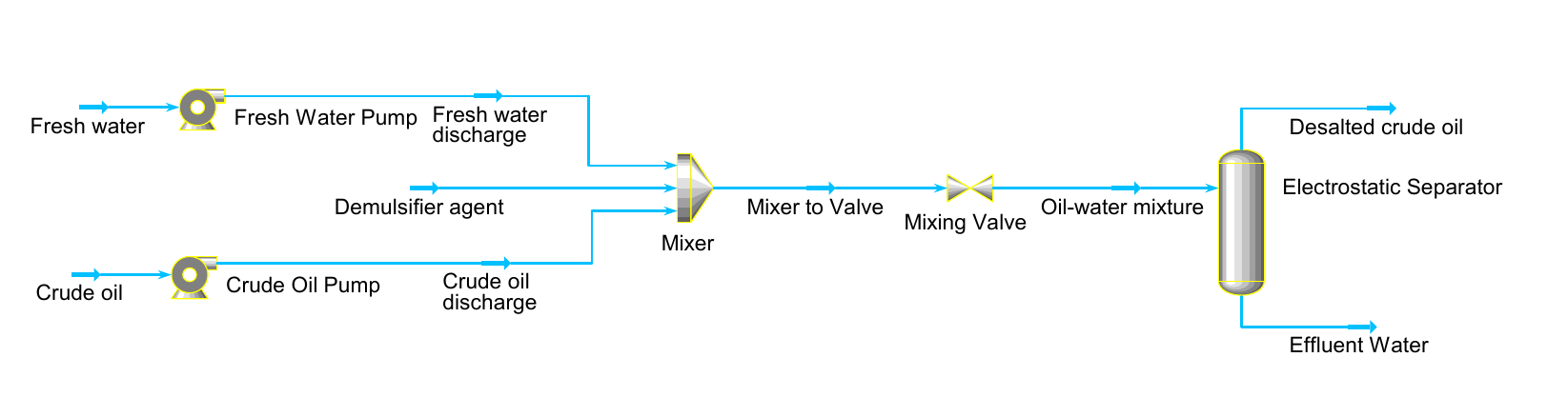}
    \caption{Generated HYSYS flowsheet corresponding to the desalting process}
    \label{fig:cs1r}
\end{figure}

For Case Study 1 (Desalting Process), the system achieved full structural consistency, correctly identifying all feed materials, unit operations, streams, and connections. The resulting simulation model, shown in Figure \ref{fig:cs1r}, executed without errors and required no fixing-loop intervention. To ensure simulator compatibility, the system introduced a mixer and an intermediate stream to formalize the mixing operation. This modification reflects simulator-driven structural normalization rather than hallucinated model content. Overall, this case demonstrates that the system can recover simple process topologies with complete accuracy and stable execution.

\begin{figure}[H]
    \centering
    \includegraphics[width=0.8\linewidth]{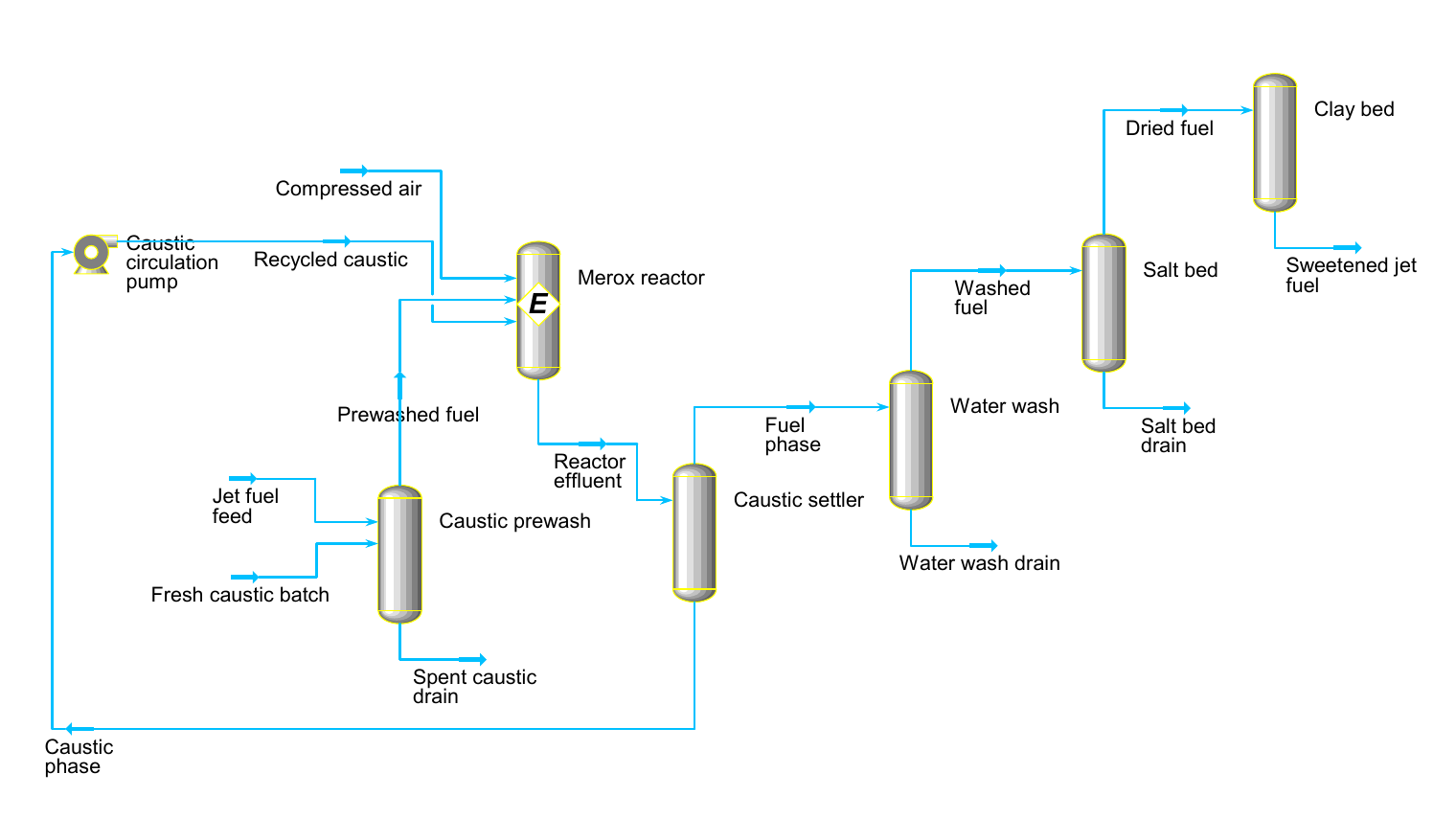}
    \caption{Generated HYSYS flowsheet corresponding to the Merox process}
    \label{fig:cs2r}
\end{figure}

\pagebreak
For Case Study 2 (Merox Process), structural consistency remained complete and the generated model executed successfully, as shown in Figure \ref{fig:cs2r}. The system correctly inferred connectivity between the caustic prewash vessel and the Merox reactor, although this connection was omitted from the diagram. This result demonstrates robust topological reconstruction under moderate visual ambiguity, showing that the system can recover process-consistent connections while preserving executability.

\begin{figure}[H]
    \centering
    \includegraphics[width=0.8\linewidth]{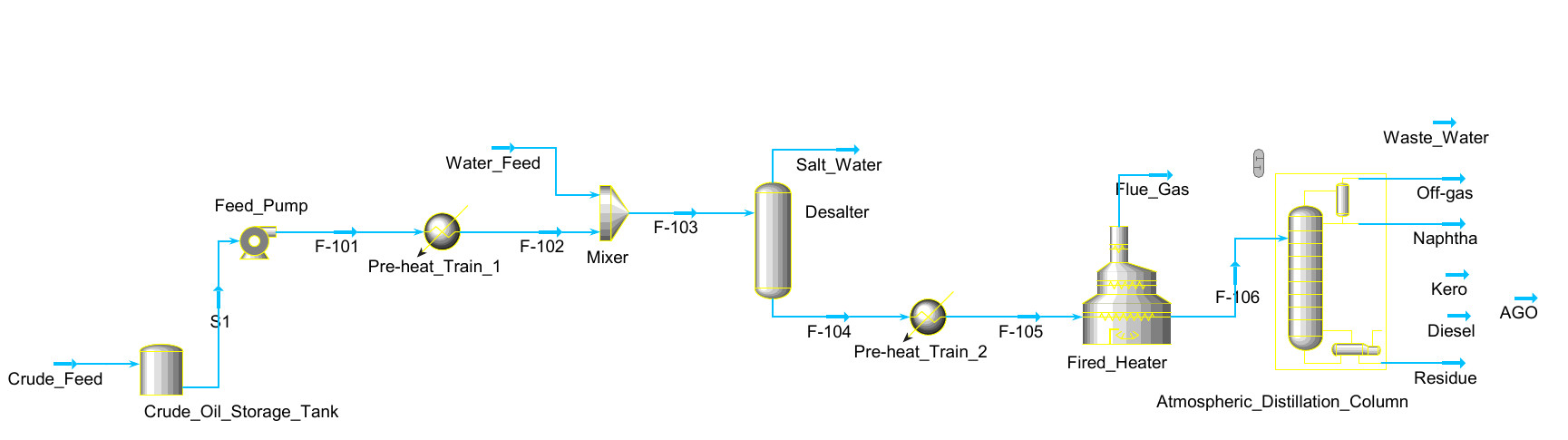}
    \caption{Generated HYSYS flowsheet corresponding to the crude distillation process}
    \label{fig:cs3r}
\end{figure}

In Case Study 3 (Atmospheric Crude Oil Distillation Process), the system remained accurate in identifying the core structural elements of the flowsheet. A slight reduction in connection consistency was observed, as shown in Figure \ref{fig:cs3r}, primarily in relation to side-draw connections from the distillation column. This discrepancy originated at the automation interface rather than the interpretation stage. As discussed in Section \ref{sec:3.2.3}, distillation columns in Aspen HYSYS are instantiated using predefined internal templates that encapsulate stage-level connectivity, thereby limiting direct programmatic control over certain side-stream attachments through the Python COM automation interface. Consequently, the deviation reflects simulator-interface limitations rather than errors in diagram interpretation, since the relevant side-draw streams were correctly represented in the graph-based Intermediate Representation.

\begin{figure}[H]
    \centering
    \includegraphics[width=1\linewidth]{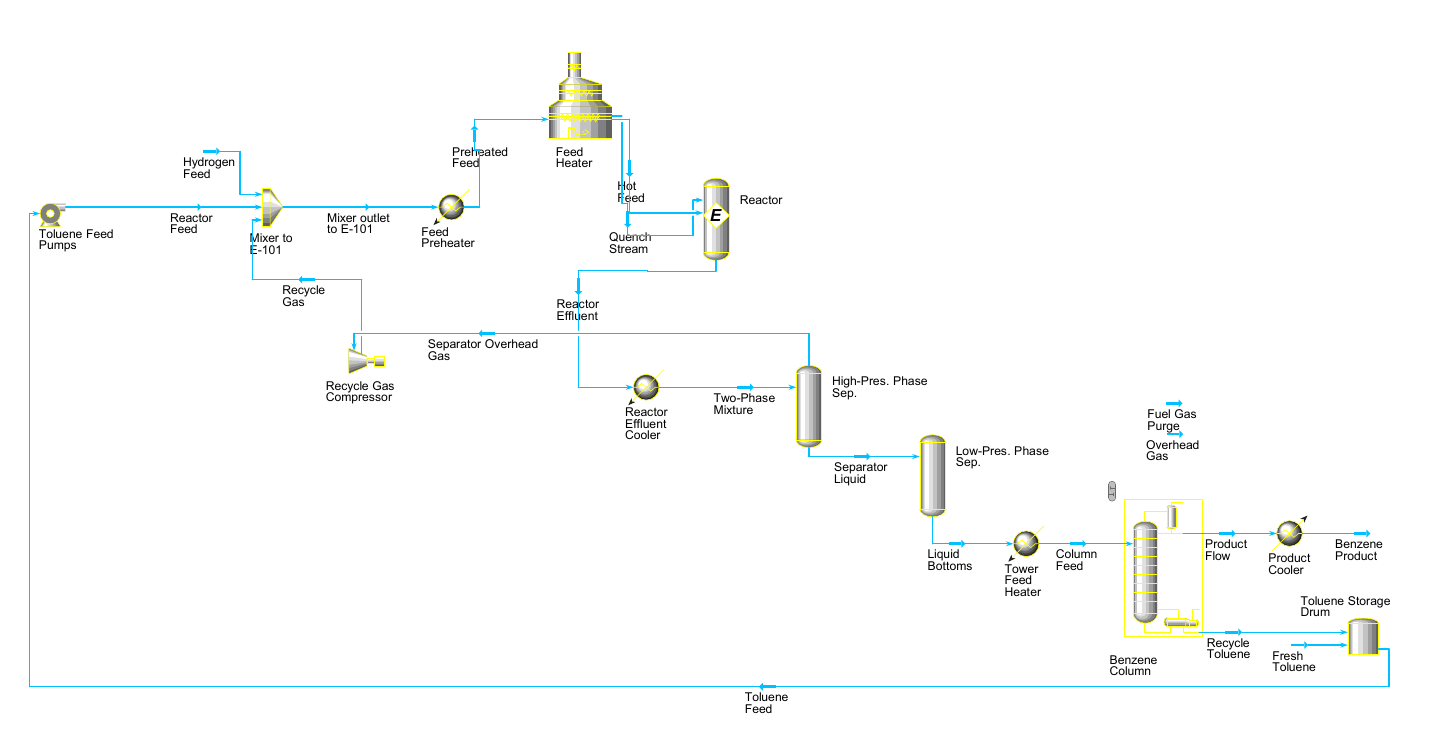}
    \caption{Generated HYSYS flowsheet corresponding to the aromatic production process}
    \label{fig:cs4r}
\end{figure}

Case Study 4 (Aromatic Production Process) represents the most complex industrial-scale flowsheet evaluated. Structural fidelity remained high (F1 $\approx 0.98$), with minor deviations in stream and connection consistency, as shown in Figure \ref{fig:cs4r}. The generated model omitted the fuel gas header and one recycle stream between the feed pump and storage drum. Additionally, the quench stream source was misassigned from the recycle gas compressor to the feed heater. These discrepancies occurred within densely interconnected recycle sections, where multiple overlapping streams increase tracing difficulty. Despite these minor structural deviations, the overall model executed successfully in HYSYS. This result indicates that the system remains highly robust even in industrial-scale flowsheets, despite minor connection errors in branched stream networks.

Overall, the results demonstrate strong structural fidelity across case studies of increasing complexity. Minor reductions in performance are primarily associated with interconnection density and simulator-specific limitations rather than errors in diagram interpretation. Execution stability remained consistent across all cases, underscoring the robustness of the multi-agent system across diverse process flow diagrams.

\subsection{Ablation Analysis}\label{sec:5.3}

Four ablation studies were designed to assess the contribution of individual components within the multi-agent workflow. As summarized in Table \ref{tab:ablation_configurations}, each configuration selectively disables or modifies a specific agent within either the diagram interpretation layer or the model synthesis layer. This setup enables a systematic assessment of how disabling a given component affects structural consistency and simulation executability relative to the full-workflow baseline (C0). The analysis is conducted across Case Study 2 (Merox Process) and Case Study 4 (Aromatic Production Process) to evaluate architectural robustness under moderate and high process complexity.

\begin{table}[H]
\centering
\caption{Ablation configurations}
\label{tab:ablation_configurations}
\begin{tabularx}{\textwidth}{>{\raggedright\arraybackslash}p{4.2cm}
                                >{\centering\arraybackslash}p{1.8cm}
                                X}
\toprule
Configuration & Target & Purpose \\
\midrule
C0 -- Full Workflow     & All    & Baseline configuration \\
C1 -- Remove Descriptor & A1     & Evaluate the contribution of visual-text grounding \\
C2 -- Remove Normalization & A3  & Evaluate the contribution of structural refinement \\
C3 -- Merge Coding Agents & B1--B3 & Evaluate the contribution of modular code decomposition \\
C4 -- Disable RAG       & B1     & Evaluate the contribution of retrieval-based material mapping \\
\bottomrule
\end{tabularx}
\end{table}

\begin{figure}[H]
    \centering
    \includegraphics[width=\linewidth]{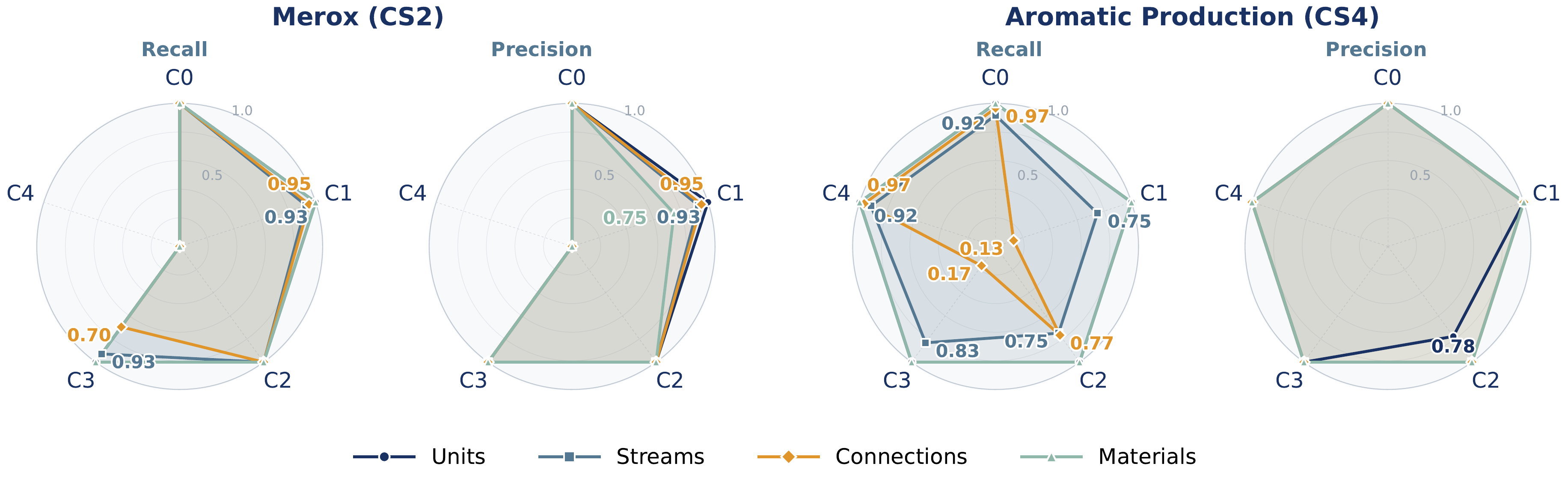}
    \caption{Ablation results across two case studies: CS2 and CS4}
    \label{fig:precision_recall}
\end{figure}

Figure \ref{fig:precision_recall} reports precision and recall for units, streams, connections, and material components across the ablation configurations (C1–C4) relative to the full-workflow baseline (C0). These plots provide a metric-level view of structural degradation under controlled architectural perturbations. Reductions in recall indicate missing structural elements, whereas declines in precision reflect the introduction of incorrect or hallucinated elements.

The ablation analysis shows that architectural sensitivity increases with process complexity. In Case Study 2 (Merox Process), unit and stream consistency remain largely preserved across configurations; however, connection consistency decreases substantially under coding consolidation (C3), indicating that modular code separation is essential for maintaining correct topology. Disabling RAG (C4) results in complete execution failure because Aspen HYSYS constructs models sequentially; unresolved material components therefore prevent successful case initialization and terminate simulation.

In Case Study 4 (Aromatic Production Process), which is characterized by dense interconnections and recycle structures, architectural modifications produce more pronounced degradation. Removal of the descriptor agent (C1) or consolidation of coding agents (C3) significantly reduces connection consistency, while omission of the normalization stage (C2) introduces hallucinated structural elements that lower precision. In contrast to Case Study 2, disabling RAG (C4) does not affect performance, as the Aromatic Production Process relies on pure components rather than mixtures. Taken together, these results indicate that structural fidelity in highly interconnected flowsheets depends critically on coordinated multi-agent processing rather than the performance of isolated agents.

\begin{figure}[H]
    \centering
    \includegraphics[width=0.9\linewidth]{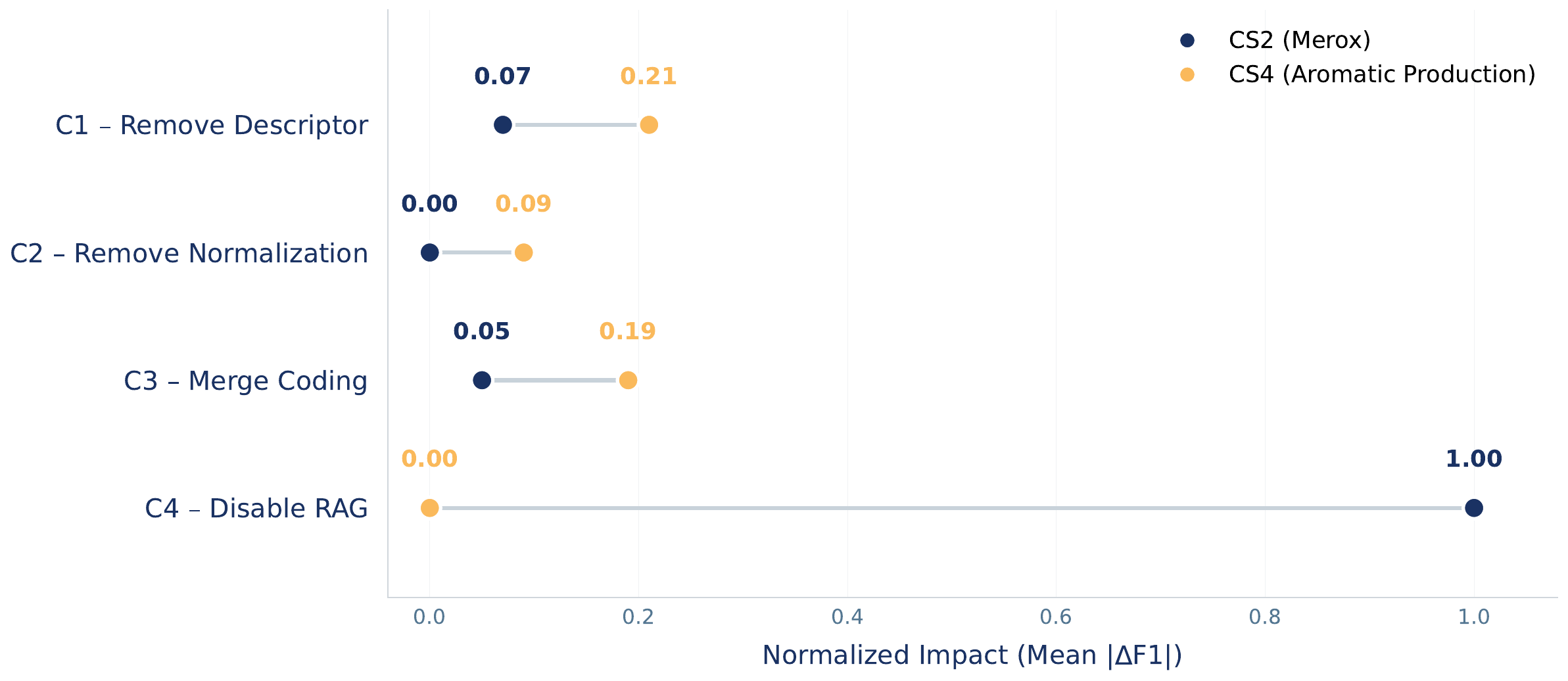}
    \caption{Normalized impact across ablation configurations}
    \label{fig:heatmap}
\end{figure}

The normalized impact dumbbell chart, shown in Figure \ref{fig:heatmap}, provides a consolidated view of the relative importance of individual workflow components. The impact score is defined as the mean absolute change in F1-score ($\Delta$F1) relative to the baseline, where $\Delta$F1 captures the deviation introduced by each ablation across units, streams, connections, and materials. Higher values, therefore, correspond to greater workflow disruption. The dumbbell chart further indicates that each component contributes meaningfully to overall robustness, with different components becoming critical as diagram complexity increases.

\subsection{Model Behavior, Robustness, and Variability}\label{sec:5.4}
To complement the previous analysis, this section examines model stability and sensitivity at the multimodal reasoning level. It focuses on reproducibility under deterministic decoding and on the influence of model architecture on connectivity reconstruction in complex process diagrams.

\subsubsection{Reproducibility Analysis}
Reproducibility was evaluated by executing the Descriptor Agent five times per case study under strictly deterministic decoding conditions, as shown in Table \ref{tab:reproducibility}. Temperature was set to 0.0, top-$k$ to 1, top-$p$ to 1.0, and the random seed was fixed at 42 to eliminate stochastic sampling. Under these settings, any observed variation reflects interpretive differences arising from visual reasoning rather than probabilistic decoding.

\begin{table}[H]
\centering
\caption{Deterministic inference parameters}
\label{tab:reproducibility}
\begin{tabularx}{\textwidth}{
    >{\raggedright\arraybackslash}p{3.0cm}
    >{\centering\arraybackslash}p{1.8cm}
    X
}
\toprule
Parameter & Value & Purpose \\
\midrule
Temperature & 0.0 & Removes probabilistic sampling \\
Top-k       & 1   & Selects only the highest-probability token \\
Top-p       & 1.0 & Disables variability from nucleus sampling \\
Seed        & 42  & Ensures consistent behavior across executions \\
\bottomrule
\end{tabularx}
\end{table}

The consistency across runs was quantified using cosine similarity between sentence embeddings of the generated descriptions, reporting both mean pairwise similarity and worst-case deviation \cite{reimers_sentence-bert_2019}. Case Studies 1, 3, and 4 exhibit near-perfect reproducibility (mean similarity $\geq 0.9889$). Case Study 2 (Merox Process) shows the only noticeable variability (mean $= 0.9594$; worst case $= 0.8986$), which is attributed to diagram-specific ambiguity rather than model instability. The Merox flowsheet contains dense textual annotations and reaction equations, increasing visual clutter, while the connection between the caustic prewash vessel and the reactor is only implicitly represented. Across trials, this ambiguous connectivity was occasionally interpreted differently, producing minor structural variation.

\subsubsection{Model Benchmark or Multimodal Architecture Benchmark}\label{sec:5.6}
A benchmarking study was conducted to evaluate the influence of multimodal model selection on reconstruction quality. The analysis focused on the interpretation layer, specifically the Descriptor Agent (A1) and Extractor Agent (A2), since their underlying multimodal models were previously identified as the primary determinants of reconstruction accuracy.
The deployed model, Gemini 3 Flash, was benchmarked against two state-of-the-art open-weight alternatives: Qwen 3-VL:235B and Qwen 3.5: 397B \cite{bai_qwen3-vl_2025, qwen_team_qwen35_2026}. The evaluation was performed on Case Study 2 (Merox Process), which was selected as a controlled stress test because of its intermediate complexity, implicit connectivity, and dense annotation, making it well-suited for assessing multimodal spatial reasoning.

\begin{figure}[H]
    \centering
    \includegraphics[width=0.43\linewidth]{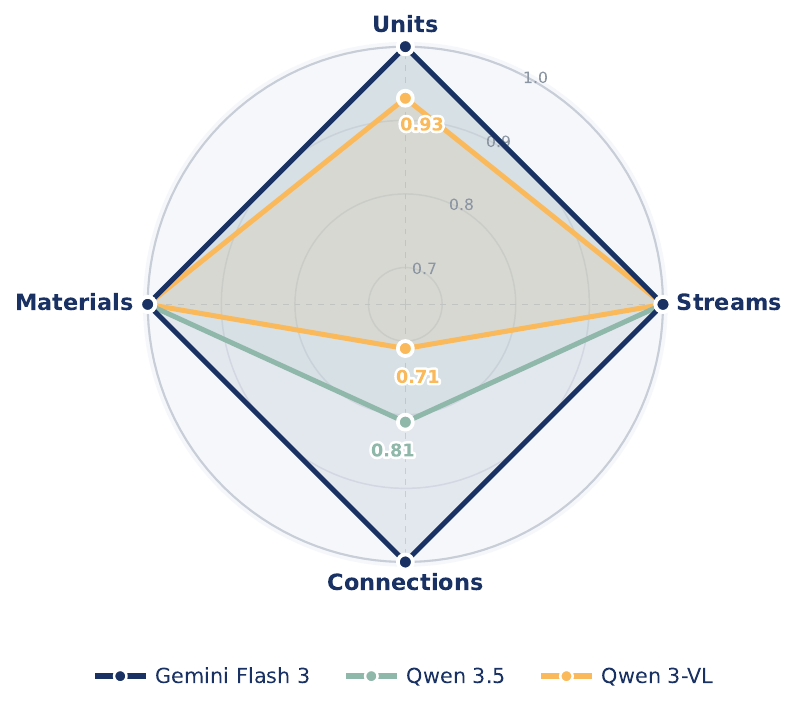}
    \caption{Comparison of F1 scores across different LLMs}
    \label{fig:model_compare}
\end{figure}

Figure \ref{fig:model_compare} summarizes the performance of the evaluated models according to the metrics defined in Section \ref{sec:5.1}. Gemini 3 Flash achieved complete consistency across all evaluated elements, accurately reconstructing units, streams, connections, and material components. Qwen 3.5 recovered the overall topology with high fidelity but exhibited connection-level inconsistencies, including misrouting of the compressed air stream and incorrect placement of liquid drain outlets. While these errors did not collapse the topology, they reduced connectivity accuracy. In contrast, Qwen 3-VL showed substantially lower robustness, hallucinating an additional equipment unit (the coalescer section) and misplacing multiple stream connections. The resulting topology required extensive manual intervention before simulation, indicating insufficient cross-modal structural alignment.

The observed performance reflects underlying architectural differences. Qwen 3-VL employs a classical late-fusion vision-language paradigm in which visual features are encoded independently before being interpreted by the language model. This separation increases susceptibility to spatial ambiguity and weakens connectivity inference. Qwen 3.5 adopts a native early-fusion multimodal architecture with improved attention mechanisms, enabling tighter cross-modal alignment and reducing connectivity errors. Gemini 3 Flash further extends multimodal integration through an iterative visual inspection mechanism, referred to as agentic vision, which enables localized refinement of ambiguous or densely annotated regions before structural commitment \cite{google_deepmind_introducing_2026}. Overall, these findings indicate that model architecture and fusion strategy are critical to object recognition and spatial reasoning in the reconstruction of complex flowsheets.

\subsection{Practical Implications and Limitations}\label{sec:5.5}
Despite strong overall performance, a few limitations were observed. These limitations can be grouped into three categories: diagram interpretation challenges, simulator constraints, and infrastructure deployment considerations.

From a visual standpoint, performance is sensitive to diagram quality and formatting. Implicit or partially drawn elements, such as units or connections, increase ambiguity and complicate process interpretation. Dense textual overlays, such as embedded reaction equations, increase OCR sensitivity and visual clutter, occasionally affecting stream tracing. In complex flowsheets, recycle loops and non-linear routing amplify small parsing deviations into measurable topological inconsistencies. Internal elements embedded within vessels, such as catalyst beds, may also be misinterpreted as standalone units in weaker multimodal models. Diagrams with omitted connections, implied operations, or under-labeled stream routing may therefore require engineering inference beyond the directly visible structure.

Simulator constraints also affect executability. Aspen HYSYS constructs models sequentially, so incorrect material component definitions can prevent case initialization and terminate execution. In addition, complex unit operations such as distillation columns rely on predefined template structures that limit dynamic stream assignment through the automation interface. As a result, the correct interpretation of the diagram does not always guarantee a directly executable simulator model because of restrictions in the simulator interface and object hierarchy.

Moreover, the system depends on carefully designed prompts, structured schemas, and simulator-specific instruction files. Consequently, transferring the multi-agent system to a different simulator or process domain may require additional adaptation. Likewise, substituting the currently deployed models may necessitate prompt refinements, as the existing prompts are tailored to the reasoning style, response behavior, and complexity-handling capabilities of those models.

From an infrastructure perspective, multimodal reasoning over high-resolution process flow diagrams is computationally intensive. Cloud-based inference introduces dependence on external compute allocation and dynamic batching, which may affect latency and reproducibility (runtime consistency). LLM inference requires increased processing time when resolving dense diagrams or performing iterative visual reasoning. Practical deployment, therefore, requires balancing structural accuracy with computational cost, response time, and hardware availability. This creates an operational trade-off in which higher reconstruction accuracy may require larger multimodal models, longer inference times, and greater hardware demand, particularly for visually dense or industrial-scale diagrams.

\section{Conclusion}\label{sec:6}
The proposed framework demonstrates the feasibility of transforming raw process diagrams into executable Aspen HYSYS simulation models through a coordinated multi-agent workflow. This claim is supported by the fact that the system successfully generated executable models in all four case studies, spanning flowsheets of increasing process and topological complexity. Structural fidelity remained perfect in the first two cases, with recovered units, streams, connections, and material components all equal to 1.00, while the more challenging Crude Distillation (CS3) and Aromatics Production (CS4) cases still maintained high performance, with only limited reductions in connection and stream consistency (CS3: CC = 0.93; CS4: SC = 0.96, CC = 0.98). In addition, the generated models remained executable even in the most complex industrial-scale case, despite dense recycle structures and minor connection deviations. By integrating multimodal extraction, structural normalization, code generation, and execution-based validation, the system therefore extends prior diagram-understanding approaches beyond descriptive reconstruction toward executable model synthesis.  

Nevertheless, the study also makes clear that the central challenge is not merely visual recognition. The more difficult problem lies in preserving engineering intent while translating imperfect, ambiguous, and sometimes implicit visual structures into forms that satisfy the rigid logical and object-level constraints of a commercial simulator. Errors arise not only from missed symbols or incorrect stream tracing, but also from mismatches between diagram conventions and simulator requirements. As a result, successful automation depends on the joint handling of perception, engineering reasoning, and simulator compatibility rather than on any one of these in isolation.

Future work should therefore focus on improving generality, robustness, and engineering realism. One important direction is to extend the framework beyond relatively clean process flow diagrams toward noisier and more heterogeneous industrial artifacts, including scanned diagrams, legacy documents, and mixed diagram-text engineering records. A second direction is to develop simulator-agnostic intermediate abstractions that can support deployment across multiple process simulation environments rather than a single commercial platform. A third direction is to strengthen self-correction through confidence-aware extraction, retrieval of engineering design rules, and more explicit simulator-in-the-loop repair cycles. Finally, broader validation on larger and more diverse industrial case studies will be needed to assess scalability, transferability, and practical deployment readiness. Taken together, these directions point toward a broader class of engineering systems in which visual interpretation, domain reasoning, and executable model synthesis are integrated into a unified automation pipeline.

\section{Data Availability \& Reproducibility}
The project codebase is publicly available in the open-source GitHub repository \url{https://github.com/OptiMaL-PSE-Lab/Sketch2Simulation}. The repository also includes the case study diagrams required for the analysis. With access to the necessary language models and a HYSYS licence, the results presented in this work can therefore be reproduced. 
Certain agent instruction files, including \texttt{instantiation\_instructions\_*.txt} and \texttt{configuration\_instructions\_*.txt}, are not included in the repository. These files contain proprietary HYSYS-domain prompt engineering materials and must therefore be obtained separately.

\section{Acknowledgments}
Financial support provided by BASF SE, EPSRC IConIC Prosperity Partnership (EP/X025292/1), and EPSRC CDT (EP/S023232/1) is acknowledged.

\printbibliography

\end{document}